\documentclass[11pt]{article}
\usepackage{graphicx}
\usepackage{epsfig}

\newcommand{\BABARPubYear}    {00}

\newcommand{\BABARProcNumber} {18}
\newcommand{\SLACPubNumber} {8692}

\def\lbabar{\mbox{{\large\sl B}\hspace{-0.4em} {\normalsize\sl A}\hspace{-0.03em}{\large\sl B}\hspace{-0.4em} {\normalsize\sl A\hspace{-0.02em}R}}}
\usepackage{relsize}
\def\babar{\mbox{\slshape B\kern-0.1em{\smaller A}\kern-0.1em
    B\kern-0.1em{\smaller A\kern-0.2em R}}}

\def\Kbar  {\kern 0.2em\overline{\kern -0.2em K}{}}

\def\Kzb   {\ensuremath{\Kbar^0}}
\def\KzKzb {\ensuremath{K^0 \kern -0.16em \Kzb}}

\def\Dbar  {\kern 0.2em\overline{\kern -0.2em D}{}}

\def\Dzb   {\ensuremath{\Dbar^0}}
\def\DzDzb {\ensuremath{D^0 {\kern -0.16em \Dzb}}}

\def\Bbar  {\kern 0.18em\overline{\kern -0.18em B}{}}

\def\Bzb   {\ensuremath{\Bbar^0}}

\def\BzBzb {\ensuremath{B^0 {\kern -0.16em \Bzb}}}

\mathchardef\Upsilon="7107
\def\Y#1S{\ensuremath{\Upsilon{(#1S)}}}

\mathchardef\Deltares="7101
\mathchardef\Xi="7104
\mathchardef\Lambda="7103
\mathchardef\Sigma="7106
\mathchardef\Omega="710A
\def\Deltabar   {\kern 0.25em\overline{\kern -0.25em \Deltares}{}}
\def\Lbar {\kern 0.2em\overline{\kern -0.2em\Lambda\kern 0.05em}\kern-0.05em{}}
\def\Sigbar{\kern 0.2em\overline{\kern -0.2em \Sigma}{}}
\def\Xibar{\kern 0.2em\overline{\kern -0.2em \Xi}{}}
\def\Obar{\kern 0.2em\overline{\kern -0.2em \Omega}{}}
\def\Nbar{\kern 0.2em\overline{\kern -0.2em N}{}}
\def\Xbar{\kern 0.2em\overline{\kern -0.2em X}{}}

\def\ev   {\ensuremath{\rm \,e\kern -0.08em V}}
\def\kev  {\ensuremath{\rm \,ke\kern -0.08em V}} 
\def\mev  {\ensuremath{\rm \,Me\kern -0.08em V}} 
\def\gev  {\ensuremath{\rm \,Ge\kern -0.08em V}} 
\def\gevc {\ensuremath{{\rm \,Ge\kern -0.08em V\!/}c}} 
\def\tev  {\ensuremath{\rm \,Te\kern -0.08em V}}
\def\mevc {\ensuremath{{\rm \,Me\kern -0.08em V\!/}c}} 
\def\gevcc{\ensuremath{{\rm \,Ge\kern -0.08em V\!/}c^2}} 
\def\mevcc{\ensuremath{{\rm \,Me\kern -0.08em V\!/}c^2}}

\def\mus  {\ensuremath{\rm \,\mus}}

\def\mus        {\ensuremath{\,\mu{\rm s}}}    
\def\gsim{{~\raise.15em\hbox{$>$}\kern-.85em
          \lower.35em\hbox{$\sim$}~}}
\def\lsim{{~\raise.15em\hbox{$<$}\kern-.85em
          \lower.35em\hbox{$\sim$}~}}

\def\pep2{PEP-II}

\newcommand{\eqref}[1]{Eq.~(\ref{eq:#1})}

\newcommand{\epjc}      [1]  {{Eur.\ Phys.\ Jour.\ C~{\bf #1}}}

\newcommand{\pl}        [1]  {{Phys.\ Lett.\ {\bf #1}}}      

\newcommand{\zp}        [1]  {{Z.\ Phys.\ {\bf #1}}}

\def\jetset74   {\mbox{\tt Jetset \hspace{-0.5em}7.\hspace{-0.2em}4}}

\long\def\inst#1{\par\nobreak\kern 4pt\nobreak
    {\it #1}\par\vskip 10pt plus 3pt minus 3pt}

\begin{document}
{\pagestyle{empty}

\begin{flushright}
SLAC-PUB-\SLACPubNumber \\
\babar-PROC-\BABARPubYear/\BABARProcNumber \\
August, 2000 \\
\end{flushright}

\par\vskip 4cm

\begin{center}
\Large \bf 
 $B$ decays to $D_s^{(*)}$ and $D^*$
\end{center}
\bigskip

\begin{center}
\large 
Gloria Vuagnin\\
Sezione INFN di Trieste\\
Area di Ricerca, Padriciano 99,  34012 Trieste, Italy\\
(for the \lbabar\ Collaboration)
\end{center}
\bigskip \bigskip

\begin{center}
\large \bf Abstract
\end{center}

The $e^+e^-$ annihilation data recorded with the \babar\ detector has
been used to study $B^0$ decays to $D_s^{(*)+}$ and $D^{*-}$ mesons. The production fraction of 
inclusive $D_s^{(*)+}$ and the corresponding momentum spectra have
been determined.
Exclusive decays $B^0 \rightarrow D^{*-}D_s^{(*)+}$ have been identified
with a partial reconstruction technique and their branching ratios have
been measured.
Fully reconstructed $B^0$ decays in the hadronic
modes $B^0 \rightarrow D^{*-} \pi^+$ and $B^0 \rightarrow D^{*-} \rho^+$
have been also studied and the measurement of their absolute branching
fractions is reported.

\vfill
\begin{center}
Contributed to the Proceedings of the 30$^{th}$ International 
Conference on High Energy Physics, \\
7/27/2000---8/2/2000, Osaka, Japan
\end{center}

\vspace{1.0cm}
\begin{center}
{\em Stanford Linear Accelerator Center, Stanford University, 
Stanford, CA 94309} \\ \vspace{0.1cm}\hrule\vspace{0.1cm}
Work supported in part by Department of Energy contract DE-AC03-76SF00515.
\end{center}

\setlength\columnsep{0.20truein}
\twocolumn
\def\sloppy{\tolerance=100000\hfuzz=\maxdimen\vfuzz=\maxdimen}
\sloppy
\vbadness=12000
\hbadness=12000
\flushbottom
\def\figurebox#1#2#3{%
        \def\arg{#3}%
        \ifx\arg\empty
        {\hfill\vbox{\hsize#2\hrule\hbox to #2{\vrule\hfill\vbox to #1{\hsize#2\vfill}\vrule}\hrule}\hfill}%
        \else
        {\hfill\epsfbox{#3}\hfill}%
        \fi}

\section{Introduction}

The study of $D_s^{(*)+}$ production in $B^0$ decays allows us to understand
the mechanisms leading to the creation of $c\bar{s}$ quark pairs.
The precise measurement of the momentum spectrum determines the fraction
of two body and multibody decay modes, and 
consequently helps to understand the $b \rightarrow c\bar{c}s$ transitions.
In this study we report a new measurement of the inclusive
$D_s^{(*)+}$ production rate in $B^0$ decays and
the branching fraction of two specific two-body 
$B^0$ decay modes
involving a $D_s^{(*)+}$ meson. 
We also have performed a study, with full reconstruction, of
the decay modes $B^0 \rightarrow
D^{*-} \pi^+$ and $B^0 \rightarrow D^{*-} \rho^+$ and measured the
corresponding branching fractions. These measurements are interesting for
testing factorization models of B decays to open charm.
Throughout this paper, conjugate modes are implied.

\section{The dataset}

The data were collected with the \babar\ detector
while operating in the PEP-II storage ring at the Stanford Linear
Accelerator Center.
 For the inclusive $D_s^{(*)+}$ production in $B^0$ decays and the 
$B^0 \rightarrow D^{*-} D_s^{(*)+}$ branching fraction measurements we used a data
sample equivalent to $7.7$ fb$^{-1}$ of integrated luminosity collected
while running on the $\Upsilon(4S)$ resonance and a sample of $1.2$ fb$^{-1}$ 
collected $40$ MeV below the $B\bar{B}$ threshold.
The measurements of the branching fractions $B^0 \rightarrow D^{*-}
\pi^+$ and $B^0 \rightarrow D^{*-} \rho^+$ use  a
subset of the same data sample corresponding to an integrated luminosity 
of $5.2$ fb$^{-1}$.

\section{Inclusive $D_s^{(*)+}$ production in $B^0$ decay}

The $D_s^{+}$ mesons are reconstructed in the decay mode $D_s^{+} \rightarrow
\phi \pi^{+}$ where $\phi \rightarrow K^+K^-$.
Particle identification is crucial to obtain a clean sample. 
Three charged tracks combining to from a common vertex are considered to be a 
$D_s^{+}$ candidate. Two of this tracks, with opposite charge, are required to be
identified  as
kaons and their invariant mass must be within $8$ MeV of the nominal
$\phi$ mass. In this decay channel, the $\phi$ meson is polarized longitudinally
which means the helicity angle of the decay, $\theta_H$  has a  cos$^2
\theta_H$ dependence\cite{conf0013}. The requirement $|cos
\theta_H|>0.3$ keeps $97.5\%$ of the signal while
rejecting $30 \%$ of the background.
The $D_s^{*+}$ are reconstructed in the decay channel $D_s^{*+}
\rightarrow D_s^{+} \gamma$ where $D_s^{+} \rightarrow \phi
\pi^{+}$. $\phi\pi^+$ combinations  within $2.5\sigma$ of the nominal $D_s^+$ mass are taken as 
 $D_s^+$ candidate. Photons must have a minimum energy of at least  $50$ MeV.
The number of $D_s^{+}$ mesons is extracted by a Gaussian fit of the 
$\phi\pi^{^+}$ invariant mass distribution for different momentum
ranges in the $\Upsilon(4S)$ rest frame.
Similarly, the number of $D_s^{*+}$ is extracted by fitting
the mass difference $m_{D_s^{*+}}-m_{D_s^+}$ distribution.    
The efficiency-corrected number of reconstructed $D_s^{+}$ as
a function of their momentum is shown in Fig.~\ref{fig:momspectra}.

In order to determine the $D_s^{(*)+}$ momentum spectrum for the
continuum, on resonance data with momentum higher than $2.45$ GeV/$c$
and off resonance data, scaled according to the luminosity ratio, have
been fitted after efficiency correction using the Peterson fragmentation 
function. The momentum spectrum of $D_s^{(*)+}$ produced in $B^0$
decays is obtained by subtracting the value of
the fit function from the on resonance data after correcting for efficiency. 
\begin{figure}
\epsfxsize160pt
\figurebox{120pt}{160pt}{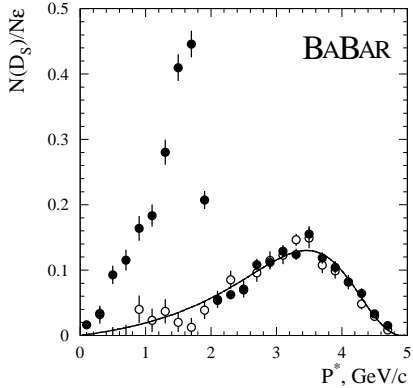}
\caption{Momentum spectrum efficiency-corrected for $D_s^{+}$. 
Solid circles indicate the on
resonance data point, while open circles are for data collected off
resonance scaled according to the luminosity. The solid line shows the
result of the fit with a Peterson
fragmentation function. }
\label{fig:momspectra}
\end{figure}
The measured branching fractions are
$(11.9\pm0.3\pm1.1)\times 10^{-2}$ and $(6.8\pm0.7\pm0.8)\times 10^{-2}$ for $B^0 \rightarrow D_s^{+} X$ 
and $B^0\rightarrow D_s^{*+} X$ respectively, assuming a $D_s^{+}\rightarrow \phi\pi^{+}$ branching fractions of $3.6\pm0.9\%$.

\section{Measurement of $B^{0} \rightarrow D^{*-}D_s^{(*)+}$ branching fractions}

The measurement of the branching fractions for the decays $B^{0} \rightarrow
D^{*-}D_s^{+}$ and $B^{0} \rightarrow D^{*-}D_s^{*+}$  uses
a partial reconstruction technique. The $D_s^{(*)+}$ are fully reconstructed, 
but no attempt is made to identify the $\bar{D}^0$ coming from the
$D^{*-}$  decay.
Instead, we combine a $D_s^{(*)+}$ candidate with a $\pi^-$ and assume 
their origin is a $B^0$ meson. We then calculate
the missing invariant mass which should be the $\bar{D}^0$ mass if our
hypothesis is correct.
The yield of $B^{0} \rightarrow D^{*-}D_s^{(*)+}$ is evaluated by fitting
the missing mass distribution (Fig.~\ref{fig:brdsdst}) with a 
 Gaussian and a background function\cite{conf0013}.
\begin{figure}
\epsfxsize160pt
\figurebox{120pt}{160pt}{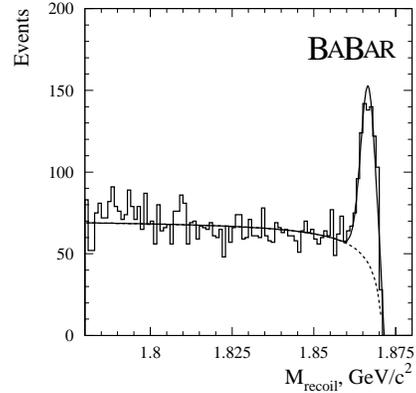}
\caption{Missing mass distributions for the $D_s^{\pm}-\pi$ systems 
before background subtraction. }
\label{fig:brdsdst}
\end{figure}
The measured branching fractions are $(7.1\pm2.4\pm2.5\pm1.8)\times
10^{-3}$ for the cannel  $B^{0} \rightarrow D^{*-}D_s^{+}$ and
for $B^{0} \rightarrow D^{*-}D_s^{*+}$ 
$(2.5\pm0.4\pm0.5\pm0.6)\times 10^{-2}$  assuming a $D_s^{+}\rightarrow \phi\pi^{+}$ branching fractions of $3.6\pm0.9\%$\cite{ref:pdg98}.


\section{Measurement of $B^{0} \rightarrow D^{*-}\pi+$ and $B^{0}
\rightarrow D^{*-}\rho^+$ branching fractions} 
 
$B^0$ candidates in the channel $D^{*-}\pi^+$ and $D^{*-}\rho^+$ are 
fully reconstructed using the decay chain 
$D^{*-} \rightarrow \bar{D^0}\pi^-$, followed by $\bar{D}^0 \rightarrow
K^+ \pi^-$. The $\rho^+$ is seen in its decay to $\pi^+\pi^0$.
The selection of events is based on a few simple criteria. Tracks are
required to originate from the beam spot and
no particle identification is used. Photons with energy greater 
than $30$ MeV are combined to form $\pi^0$ candidates. Kaons and pions 
with opposite charge and  coming from the same vertex
must have an invariant mass within $\pm 2.5\sigma$ of the nominal $D^0$ mass
to form a  $D^0$ candidate.
The ${D}^0$ candidates are required to have a momentum greater than
$1.3$ GeV/$c$ in the $\Upsilon(4S)$ frame and are combined with a pion to
form a charged $D^{*}$ candidate.
We require 
$\Delta m = m(\bar{D}^0\pi^-)-m(\bar{D}^0)$ to be within $2.5\sigma$ of 
the nominal mass difference $D^{*-}-\bar{D}^0$. The  $D^{*-}$ is combine with a $\pi^+$ candidate,
 with a momentum greater than $500$ MeV/$c$ or a $\rho^+$ to form 
$B^0$ candidates.
 In the decay $B^{0} \rightarrow D^{*-}\pi^+$ the longitudinal polarization 
 of the  $D^{*-}$ is used to reduce background\cite{conf0006}.
 For the $B^{0} \rightarrow D^{*-}\rho^+$ mode, $\rho^+$ candidate are
 selected requiring the $\pi^+\pi^0$ invariant mass  within
 $150$ MeV/$c^2$ of the $\rho^+$ nominal mass. Event shape variables are also
 used to remove continuum background.

For correctly reconstructed $B^0$ mesons, the energy of the $B^0$ candidate, 
$E^*_{B^0}$ must be equal to the beam energy $E^*_b$ were both are 
evaluated at the 
$\Upsilon(4S)$ frame. We define $\Delta E =E^*_{B^0}-E^*_b $. 
The beam energy substituted mass, $m_{ES}$ is defined as
$  m_{ES}^2  =  \left(E^*_b\right)^2 -\left(\sum_i \mbox{\boldmath 
$p$}_i\right)^2$, where  \mbox{\boldmath$p$}$_i$ is the
 momentum of the $i$th daughter of the $B$ candidate.
  The variables $\Delta E$ and $m_{ES}$ are used to define the signal
  and sideband regions.
 For both modes, the region between $5.2$ and $5.3$ GeV/$c^2$  in
 $m_{ES}$ and between  $\pm300$ MeV in $\Delta E$ is used to study
 signal and background properties. 
 By staying below $|\Delta E
 |=m_{\pi}$, we avoid correlated background from $B$ decays where a real
 final state pion is either not included in the reconstruction or a
 random one is added to the observed state.

 The measurement of branching fractions requires an estimate of the number
 of signal events.
A  Gaussian distribution and a background function\cite{ref:ARGUS}, which
 parametrize how the phase space approach zero as the energy approaches
 $E^*_b$, are used to fit the $m_{ES}$ distribution obtained requiring
 $|\Delta E  |< 2.5\sigma_{\Delta E}$ as shown in
 Fig.~\ref{fig:brdspi}.
\begin{figure}
\epsfxsize160pt
\figurebox{120pt}{160pt}{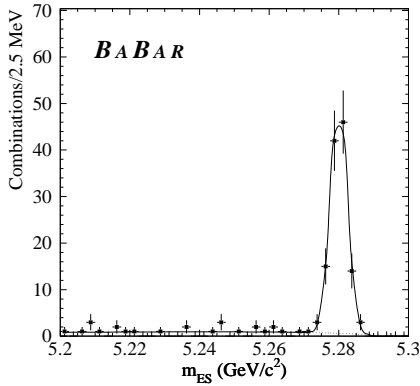}
\caption{Distribution of $m_{ES}$ for $|\Delta
 E  |< 2.5\sigma_{\Delta E}$ for the cannel 
 $B^{0} \rightarrow D^{*-}\pi^+$.  }
\label{fig:brdspi}
\end{figure}
Based on the fitted yield of signal events
the preliminary results for the branching fractions for 
$B^{0} \rightarrow D^{*-}\pi^+$ and $B^{0} \rightarrow D^{*-}\rho^+$
are $(2.9\pm 0.3\pm 0.3)\times 10^{-3}$ and
$(11.2\pm 1.1\pm 2.5)\times 10^{-3}$ respectively. 
The branching fraction for $B^{0} \rightarrow D^{*-}\rho^+$ includes all 
non-resonant and quasi-two-body contributions that lead to a $\pi^+
 \pi^0$ invariant mass in the $\rho$ band. 
However, the acceptance for non-resonant $D^{*-}\pi^+\pi^0$ decays
is about 15\% of $D^{*+}\rho^+$ so that, combined with the known
 branching fraction for this mode, the non-resonant contribution to our result
is expected to be quite small. 
Both branching fraction results compare well with previous measurements
 and with the world average\cite{ref:pdg98}.


\begin{thebibliography}{99}

\bibitem{conf0013}
\babar\ collaboration, 
"Study of inclusive $D_s^{(*)\pm}$ production in $B$ decays and
measurement of $B^0 \rightarrow D^{*-}D_s^{(*)+}$ decays using a partial
reconstruction technique", 
\babar-CONF-00/13, SLAC-PUB-8535.

\bibitem{ref:pdg98}
C.\ Caso {\em et al.}, \epjc{3} (1998) 1.

\bibitem{conf0006}
\babar\ collaboration, 
"Measurement of the branching fractions of $B^0 \rightarrow D^{*-}\pi^+$
and $B^0 \rightarrow D^{*-}\rho^+$", 
\babar-CONF-00/06, SLAC-PUB-8528.

\bibitem{ref:ARGUS}
ARGUS Collaboration, H.\ Albrecht {\em et al.}, \zp{C48} (1990) 543;
superseded results in {\em op cit.}, \pl{B185} (1987) 218; \pl{B182}
(1986) 95.

\end{thebibliography}
\end{document}